\documentclass[a4paper]{PoS}

\title{Measurements of the top quark properties at decay with CMS}

\ShortTitle{Measurements of the top quark properties at decay with CMS}

\author{\speaker{Andrea Castro}%
        \thanks{The measurements described here are presented on behalf of the CMS Collaboration.}\\
       University of Bologna and INFN - Italy\\
       E-mail: \email{Andrea.Castro@cern.ch}}

\abstract{CMS measurements of  properties related to top quark decays are discussed. The results presented regard the measurement of the W boson helicity, the investigation of  anomalous couplings in the Wtb vertex, and the search for very rare decays, such as ${\rm t \to Zq}$ and ${\rm t\to Hq}$, which are associated to  flavor-changing neutral currents.}

\FullConference{The European Physical Society Conference on High Energy Physics\\
		 5-12 July\\
		 Venice, Italy}

\begin{document}

\section{Introduction}
Because of its large mass and short lifetime, the top quark plays a special role in the standard model (SM) of particle physics. For this reason, measurements of top quark properties are important to assess the consistency of the SM, and to look for deviations from it which would be hints of physics beyond the SM (BSM). The top quark decays before hadronizing,  so certain of its properties, such as spin or polarization, 
%directly 
pass to its decay products characterizing the associated Wtb vertex.
\par
In this report, these properties ''at decay'' are measured  using data from pp collisions at the LHC collected by the CMS detector \cite{cmsdetector} in 2012 at $\sqrt{s}=8$ TeV, and corresponding to $19.7$ fb$^{-1}$. The properties discussed here refer to the W boson helicity, the investigation of  anomalous couplings in the Wtb vertex, and the search for very rare decays, namely ${\rm t \to Zq}$ and ${\rm t\to Hq}$, which are associated to  flavor-changing neutral currents (FCNC). Results are compared with the SM predictions probing for physics BSM, and when deviations are not found exclusion limits are set.

\section{Measurement of the W helicity fractions}
The W boson from the top quark decay can be produced with a left-, right- or longitudinal polarization, with the corresponding helicity fractions $F_L$, $F_R$ and $F_0$ defined as the ratio of the individual widths to the total width, $F_{L,R,0}=\Gamma_{L,R,0}/\Gamma$. These helicity fractions are sensitive to the Wtb vertex structure, and in the SM assume the values $F_0=0.687\pm 0.005$, $F_L=0.311\pm 0.005$ and $F_R=0.0017\pm 0.0001$ \cite{theory-1}, at next-to-next-to-leading order, for a top quark mass of $172.8\pm 1.3$ GeV.  These fractions are measured recurring to the distributions of the helicity angle $\theta^\star$ which is defined as the angle between the charged lepton (or down-type quark) and the reversed direction of the top quark, both measured in the W boson rest frame. 
\par
\begin{figure}[htb]
\begin{tabular}{cccc}
   \begin{minipage}{0.4\textwidth}
      \includegraphics[width=6.5cm]{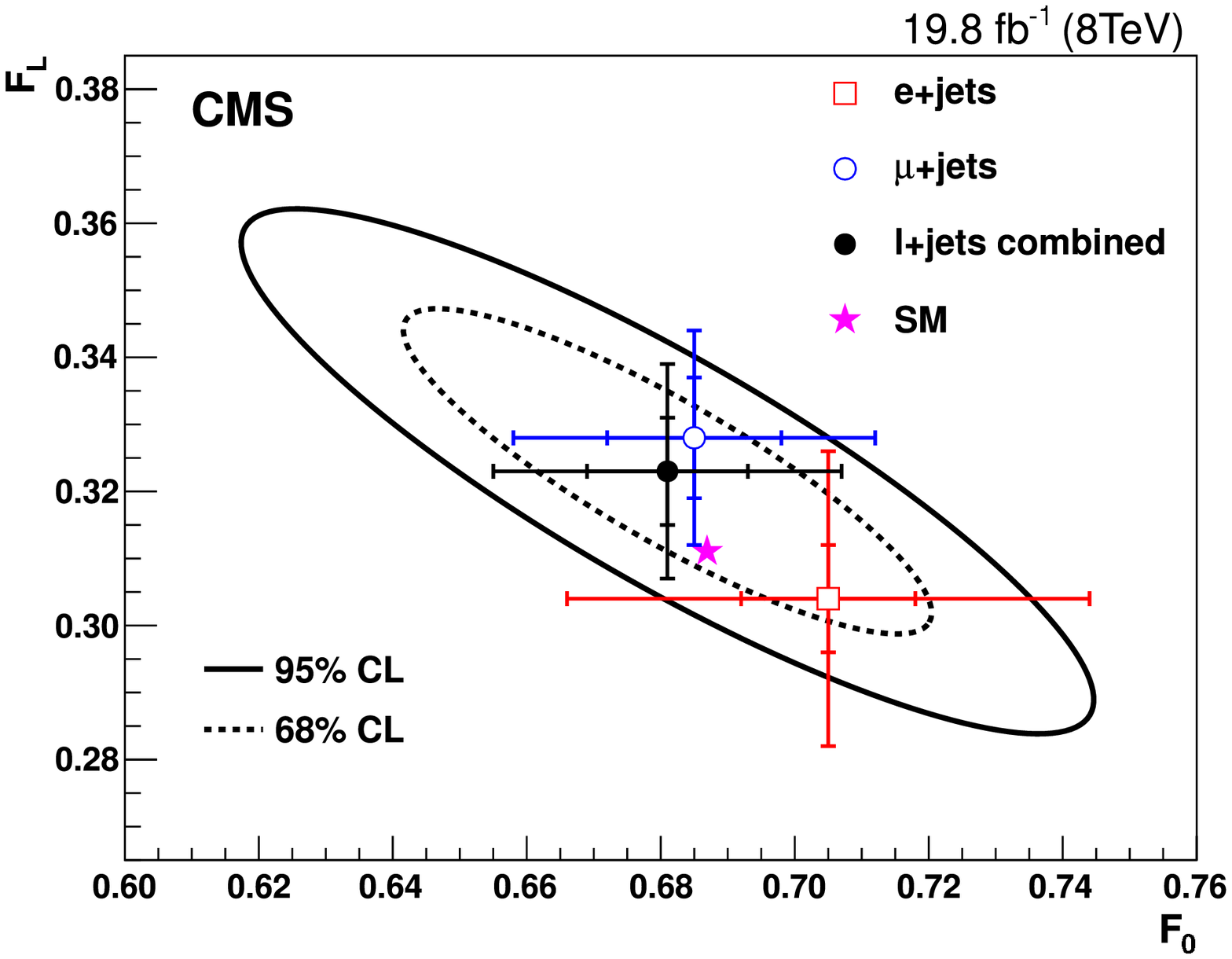}
   \end{minipage}
  & ~~~ & ~~~ &
   \begin{minipage}{0.4\textwidth} 
      \includegraphics[width=6.5cm]{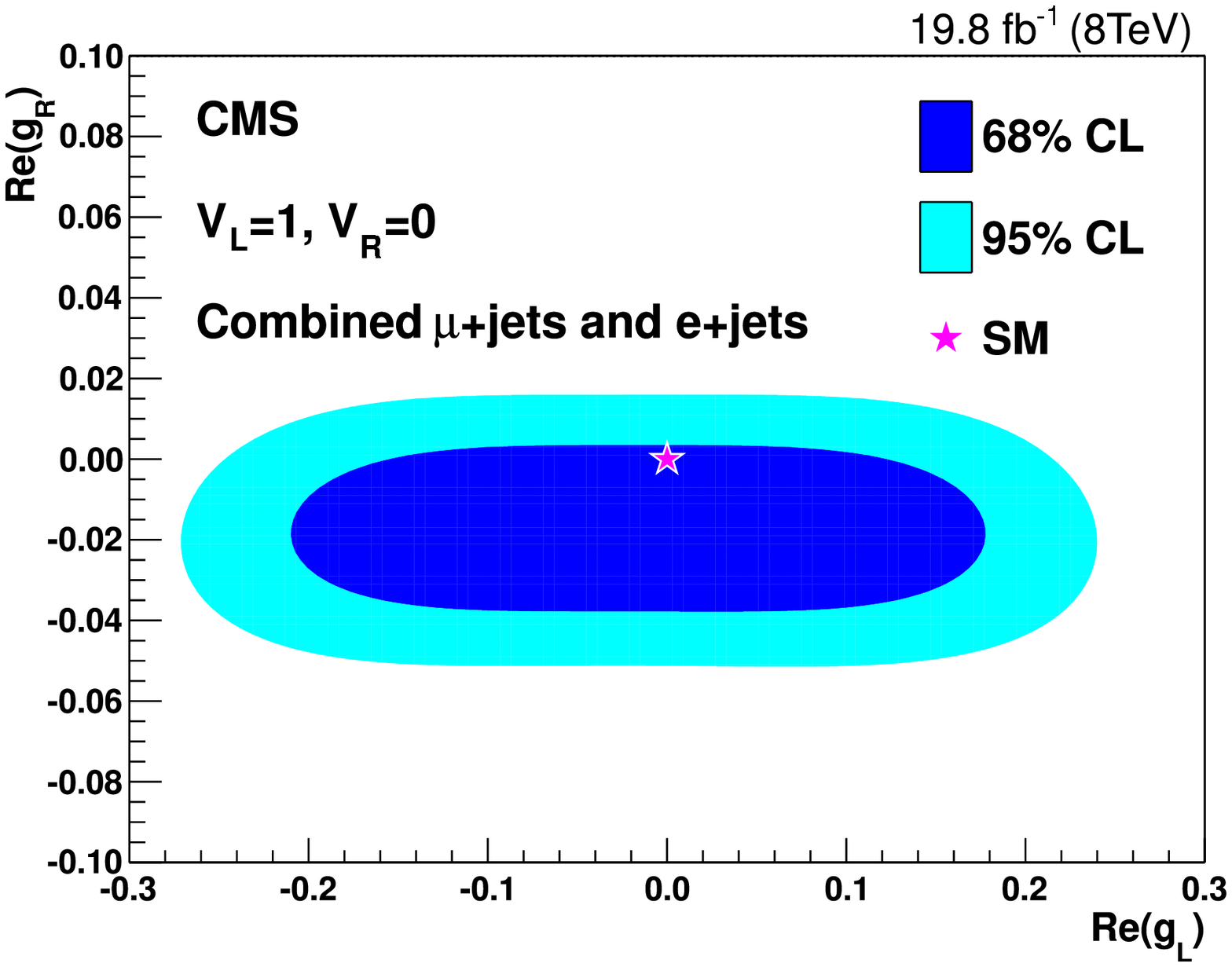}
   \end{minipage}
\end{tabular}
\caption{ Contour plots for  $\mathrm{ t\bar t}$ events~\cite{cms-1}. Left: helicity fractions in the $(F_0,F_L)$ plane. Right: anomalous couplings in the  $(Re(g_L),Re(g_R))$ plane. }
\label{helicity-fig1}
\end{figure}

This measurement is conducted on ${\rm t\bar t}$ events decaying to lepton + jets final states~\cite{cms-1}, and on single top quark events with one electron or muon plus jets~\cite{cms-2}. The event selection in the ${\rm t\bar t}$ case requires one isolated electron or muon, together with at least four jets, two of which identified as coming from b quark fragmentation. For single top quark events, one isolated electron or muon is requested, along with exactly two jets, one of which associated to a b quark; such a selection is intended to favor the so-called $t$-channel production of single top quarks.
The $\cos \theta^\star$ distributions of the selected events are compared to those expected from Monte Carlo top quark events weighted as to reproduce different values for the helicity fractions,  plus those from background events. The W boson helicity fractions are measured maximising a binned Poisson likelihood function derived from such distributions. For ${\rm t\bar t}$ events, only the leptonic branch (i.e. the one including the W boson which decays leptonically) is considered, because the hadronic branch (i.e. the one including the W boson which decays into quarks) carries an ambiguity associated to the inability to distinguish up-type from down-type initiated jets. Different channels (with an electron or muon) are first fit separately to obtain the helicity fractions $F_L$ and $F_0$ imposing the unitarity constraint $F_L+F_0+F_R=1$, then the results are combined (using the BLUE method~\cite{blue}) to account for correlations. The results of the fits for the two channels are
$F_0=0.681\pm 0.012(\mathrm{ stat})\pm 0.023(\mathrm{ syst})$,  $F_L=0.323\pm 0.008(\mathrm{ stat})\pm 0.014(\mathrm{ syst})$, and 
 $F_0=0.720\pm 0.039(\mathrm{ stat})\pm 0.037(\mathrm{ syst})$,  $F_L=0.298\pm 0.028(\mathrm{ stat})\pm 0.032(\mathrm{ syst})$, respectively.
 Figure\,\ref{helicity-fig1}-left shows the helicity measurements for  the ${\rm t\bar t}$ channel as contour plots. 
A similar plot has been obtained for single top quark events.
For both channels the measured helicities $F_L$ and $F_0$ agree with SM expectations.
\par
The structure of the Wtb vertex can be expressed with a Lagrangian function containing anomalous vector-right ($V_R$) and tensor couplings ($g_L$ and $g_R$), for which the SM expectation is $V_R=g_L=g_R=0$. Values different from zero are then an indication of deviations from the SM.
By fixing the two vector couplings to their SM expectation ($V_L=1$, $V_R=0$), limits are derived on the tensor couplings, using the theoretical relations between the helicity fractions and the couplings~\cite{theory-2}. The values allowed for ${\rm Re}(g_L)$ and ${\rm Re}(g_R)$ are shown as contour plots in Fig.\,\ref{helicity-fig1}-right for $\mathrm{ t\bar t}$ events. A similar plot has been obtained for single top quark events. For both channels the measurements are consistent with the SM prediction  $g_L=g_R=0$.

\par

\par
\section{Search for anomalous couplings}
While the V--A structure of the charged current involving light quarks is well established, the structure of the Wtb vertex is less constrained. Furthermore, FCNC interactions associated to tcg and tug vertices are predicted to be very small in the SM, but can be strongly enhanced in extensions of the SM. 
\par
Anomalous couplings in the Wtb vertex  and FCNC interactions regarding tug and tcg vertices, have been searched for by CMS in single top quark events with one  muon plus jets~\cite{cms-3}, including also data collected at $\sqrt{s}=7$  and corresponding to $5.0$ fb$^{-1}$. The selection requires one isolated muon, one jet in the forward region, one b-tagged jet and an additional softer b-jet, and aims at selecting events in the so--called $t$-channel single top quark production.  
\par
\begin{figure}[htb]
\begin{tabular}{cccc}
   \begin{minipage}{0.4\textwidth}
      \includegraphics[width=6.8cm]{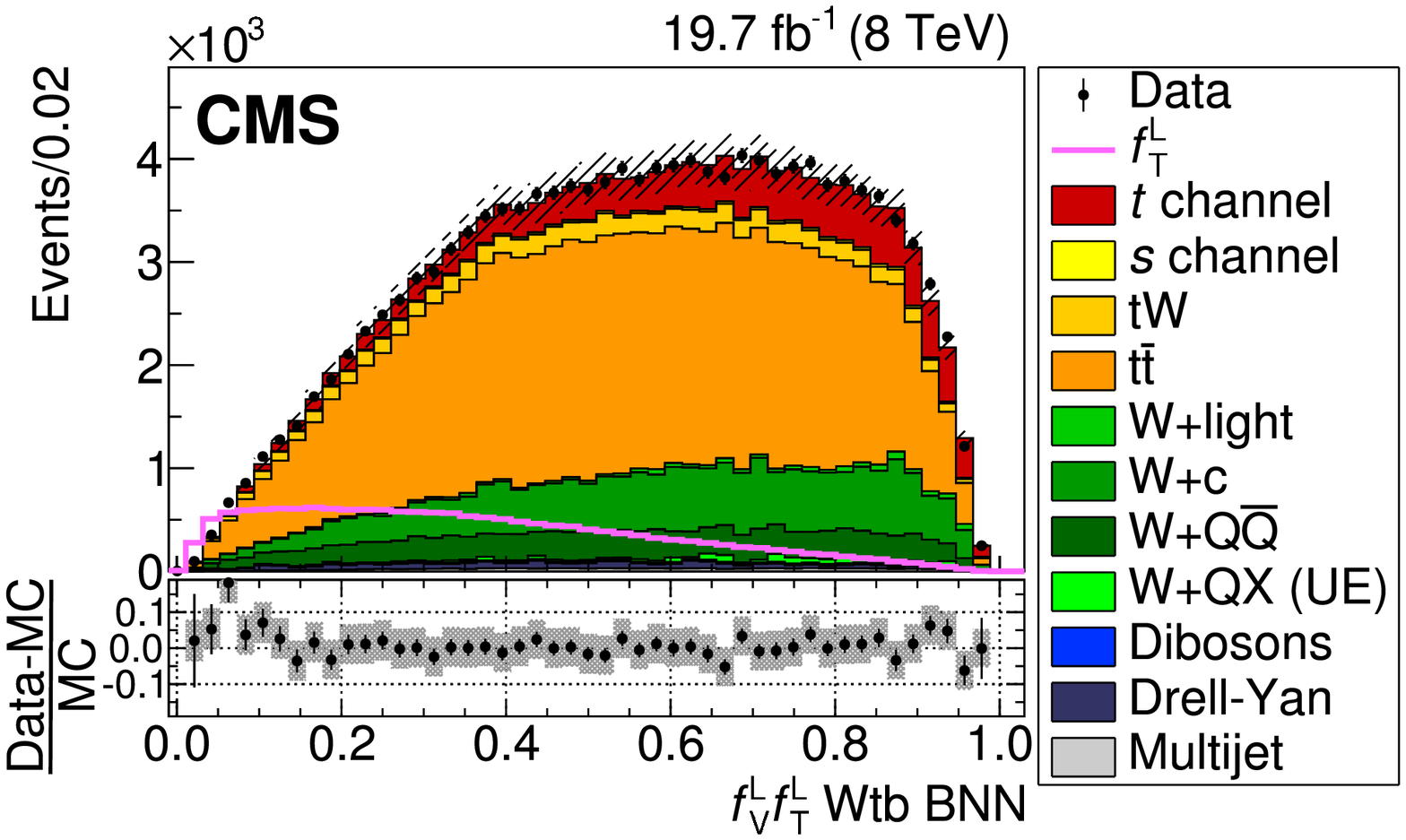}
   \end{minipage}
  & ~~~ & ~~~ &
   \begin{minipage}{0.4\textwidth} 
      \includegraphics[width=6.8cm]{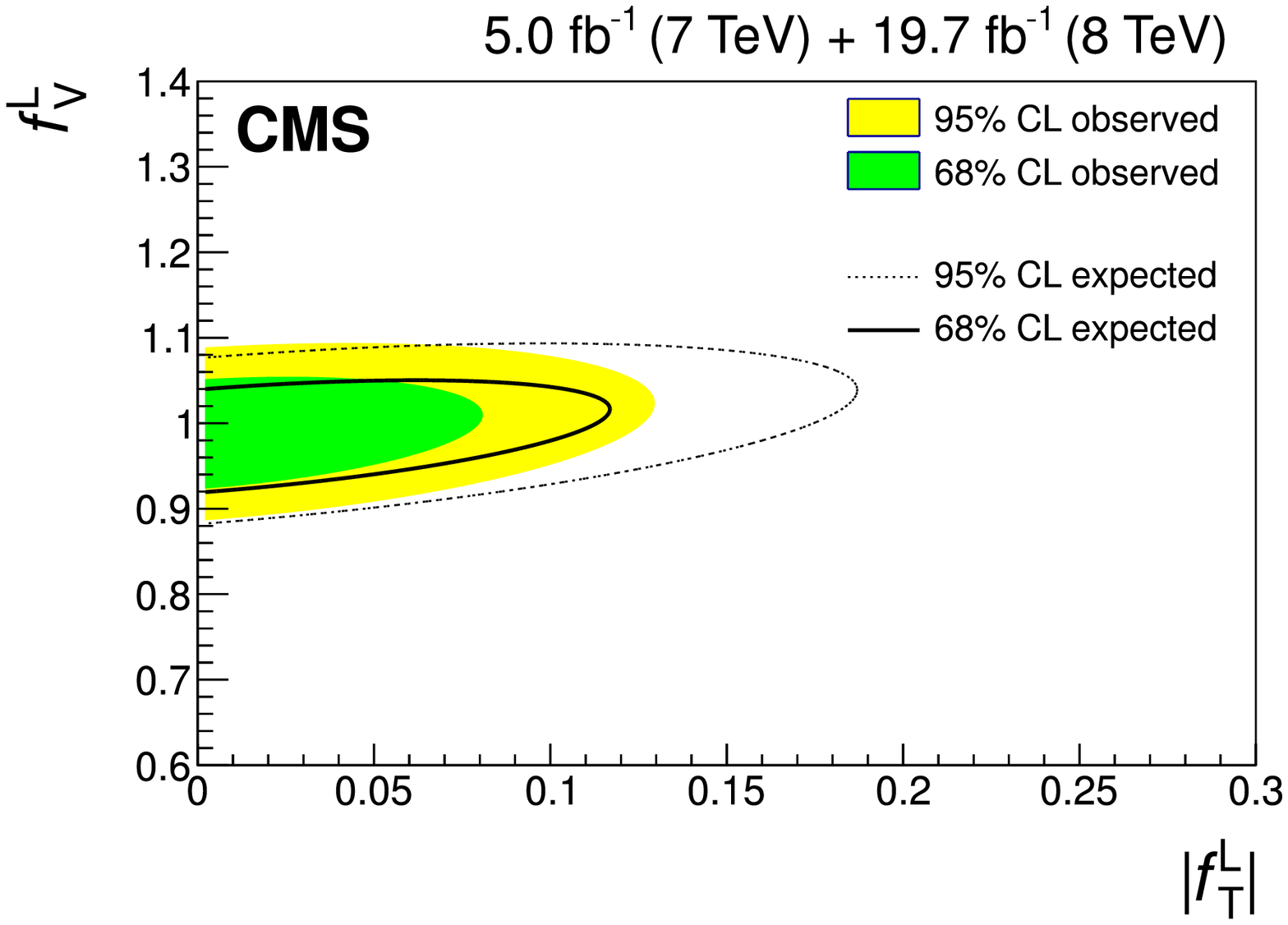}
   \end{minipage}
\end{tabular}
\caption{Single top quark events~\cite{cms-3}. Left: distribution of the Wtb discriminant. Right: exclusion limits in the $(f_V^L,\vert f_T^L\vert)$ plane. }
\label{anomalous-figs}
\end{figure}

To distinguish the $t$-channel single top quark production processes from multijet background and other SM processes, two Bayesian neural networks (BNNs) are trained and utilized. Then, three additional Wtb  BNNs are used to separate separate the individual contributions of right-handed vector $(f_V^R)$, left-handed $(f_T^L)$ and right-handed $(f_T^R)$ tensor couplings from the left-handed vector coupling $(f_V^L)$ as expected in the SM. In addition, the tcg and tug contribution are distinguished from SM processes with the help of two BNNs trained specifically for this purpose.
From the simultaneous fit of the distributions for the SM BNN and for the additional BNNs associated to the anomalous couplings, exclusion limits are set as two-dimensional contours. Fig.\,\ref{anomalous-figs}-left shows the distribution of the Wtb BNN discriminant for the scenario $(f_V^L,f_T^L)$. No $f_T^L$ contribution is seen, and exclusion limits are set, as shown in the two-dimensional contour plot of Fig.\,\ref{anomalous-figs}-right. Similar plots and limits are obtained for the other possible scenarios $(f^L_V,f^R_V)$ and $(f^L_V,f^R_T)$.

Deviations from the SM expectations can be associated to FCNC processes, whose production cross section depends on the strengths of these strong interactions, i.e. $\kappa_{\mathrm{tcg}}/\Lambda$ and $\kappa_{\mathrm{tug}}/\Lambda$, $\Lambda$ being the scale of new physics ($\Lambda\approx 1$ TeV).  New BNNs are trained to discriminate each FCNC process from  SM events. 
%Fig.\,\ref{fcnc}-left shows the distribution of the FCC BNN discriminant for the tcg coupling, with a similar plot for the tug BNN. No tcg/tug contribution is seen and exclusion limits are set, as shown in the two-dimensional contour plot of Fig.\,\ref{fcnc}-right. 
Individual 95\% confidence level (CL) observed (expected) exclusion limits on  $\kappa_{\mathrm{tug}}/\Lambda$ are obtained by integrating over 
$\kappa_{\mathrm{tcg}}/\Lambda$, and vice versa, yielding $\vert\kappa_{\mathrm{tug}}\vert/\Lambda<4.1 ~(4.8) ~\times 10^{-3}$ TeV$^{-1}$ and
 $\vert\kappa_{\mathrm{tcg}}\vert/\Lambda<1.8 ~(1.5) ~\times 10^{-2}$ TeV$^{-1}$. 
From these individual limits, upper limits on the branching fractions ${\cal B}({\rm t\to ug})$ and ${\cal B}({\rm t\to cg})$ are derived, that is ${\cal B}({\rm t\to ug})<2.0~ (2.8)~ \times 10^{-5}$ and  ${\cal B}({\rm t\to cg})<4.1~ (2.8)~ \times 10^{-4}$.

\section{Search for rare decays}
The FCNC decay of a top quark  is forbidden at tree level in SM and strongly suppressed by the GIM mechanism at higher orders. For this reason the branching fraction of the decays of a top quark to a Z boson or a Higgs boson plus an up-type quark is expected to be ${\cal O}(10^{-15}\div 10^{-14})$. However, in BSM models these fractions can become as high as ${\cal O}(10^{-5}\div 10^{-3})$ and become within reach.
\par
\begin{figure}[htb]
\begin{tabular}{cccc}
   \begin{minipage}{0.4\textwidth}
      \includegraphics[width=5.5cm]{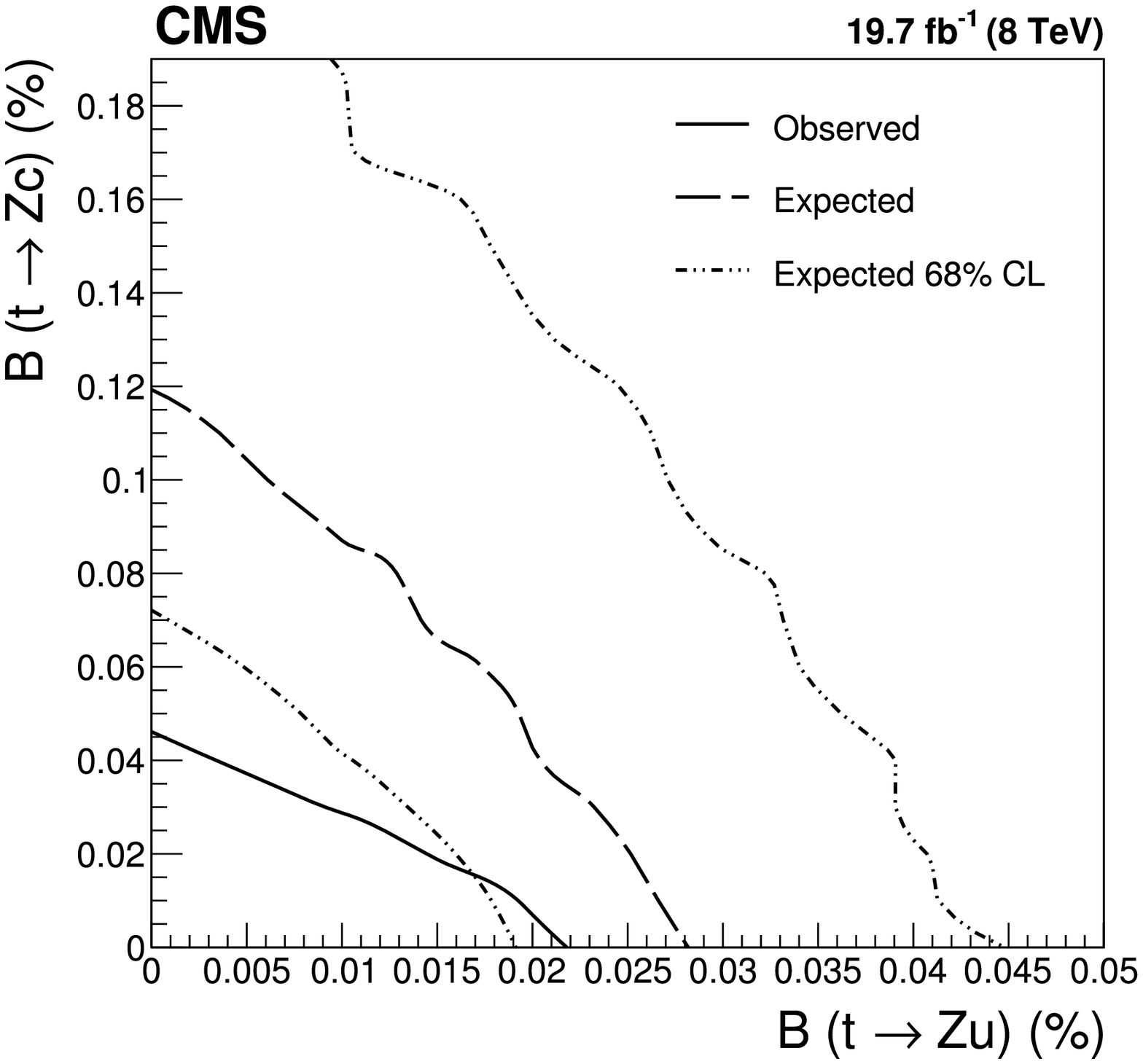}
   \end{minipage}
  & ~~~ & ~~~ &
   \begin{minipage}{0.4\textwidth} 
      \includegraphics[width=5.5cm]{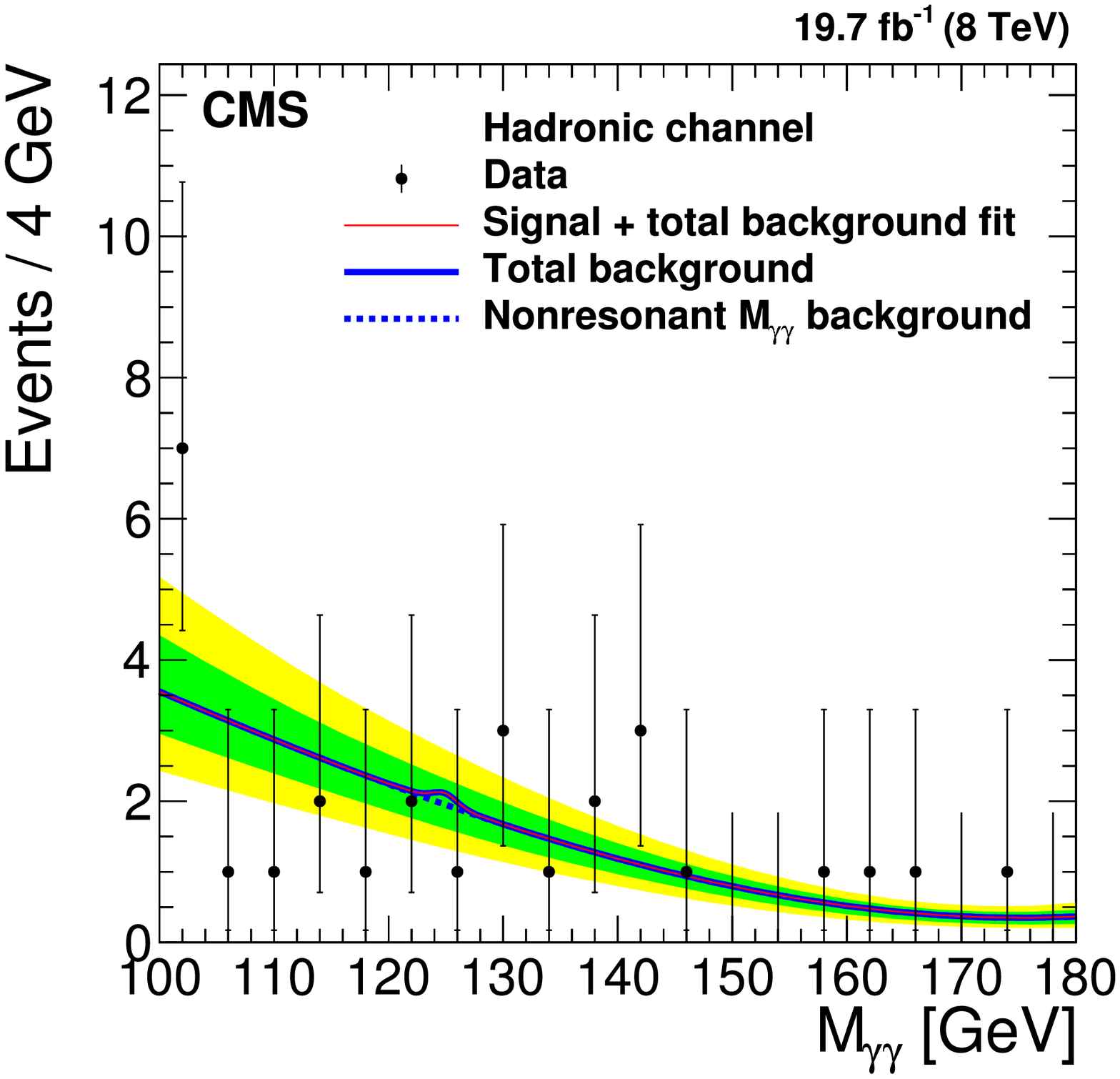}
   \end{minipage}
\end{tabular}
\caption{ Left: exclusion limits at 95\% CL on  ${\cal B}({\rm t\to Zc})$ vs  ${\cal B}({\rm t\to Zu})$\,\cite{cms-4}. Right: diphoton invariant mass for the hadronic channel~\cite{cms-5}.}
\label{rare-figs}
\end{figure}

The rare decays ${\rm t\to Zq}$ or ${\rm t\to Zc}$ are searched for in ${\rm t\bar t}$ final states  with three leptons (e or $\mu$), as described in Ref.\,\cite{cms-4}.
 These events need to be disentangled from SM production and the FCNC production of tZ events. 
The separation of the three classes of events is performed recurring to boosted decision tree (BDT) discriminants specifically trained. 
Distributions of these three discriminants are fitted to the data.
 In the case of the tZ-FCNC discriminant, the fit is consistent with the SM-only hypothesis, and limits are derived on the branching fractions  ${\cal B}({\rm t\to Zu})<0.022\%$ and  ${\cal B}({\rm t\to Zc})<0.049\%$, as shown in Fig.\,\ref{rare-figs}-left.
\par
The rare decays ${\rm t\to Hu}$ or ${\rm t\to Hc}$ are searched in ${\rm t\bar t\to Wb Hq}$ events~\cite{cms-5}, where the W boson can 
decay either leptonically or hadronically, while the Higgs boson can decay to two photons (very clean signal but with low yield), 
or to two W/Z bosons, or even to ${\rm b\bar b}$, producing different final states.
 Three independent analyses/selections are then combined: 1. multilepton analysis, requiring two same-sign leptons, or 3 leptons (e or $\mu$), aimed for ${\rm H\to WW,~ ZZ,~ \tau\tau}$; 
2. diphoton + W (leptonic or hadronic) + b-tagged jet, for ${\rm H}\to\gamma\gamma$; 
3. 3 b-tagged jets + leptonic W + additional jet, for ${\rm H\to b\bar b}$. 
The diphoton selection is the most sensitive, and the ${\rm t\to Hq, ~ H\to\gamma\gamma}$ decays are searched for by looking at the diphoton invariant mass which is fitted to look for a diphoton resonant signal in excess of the non-resonant diphoton background (plus the SM resonant background).
 No signal is seen, neither in the leptonic channel nor in the hadronic one, see Fig.\,\ref{rare-figs}-right.
No signal is observed also in the other two selections, yielding observed (expected) upper limits on the branching fractions for these rare decays, namely ${\cal B}({\rm t\to Hc)<0.40~ (0.43)\%}$ and  ${\cal B}({\rm t\to Hu)<0.55~ (0.40)\%}$ .

\section{Conclusions}
Several properties of the top quark decay have been discussed here, in order to confirm the consistency of the SM or to look for deviations from it. The W boson helicity fractions have been measured and found consistent with the SM, with no indication of anomalous couplings for the top quark. Rare decays of the top quark into Zq or Hq have been searched for with a null result. For what concerns top quark properties, the  standard model is once again confirmed.

\end{document}